Comment on "Why is the DNA Transition First Order?" and "Griffiths Singularities in Unbinding of Strongly Disordered Polymers"

The papers [1,2] consider unbinding of a disordered heteropolymer. They find the first order phase transition [1] when disorder is strong (i.e. the ratio v of the binding energies is large; in DNA v is~1.1) and the Griffiths singularity, i.e. infinite order transition, otherwise [2]. The problem is important, since many physical phenomena map onto the same model (see refs. in [1, 2]). Drastic non-universality in the strength of disorder is unanticipated. Unfortunately, both titles are misleading: they claim the results which are proven for homopolymers only.

Kafri and Mukamel (KM, [2]) consider infinite v, they assume that the larger binding energy is infinite. This invalidates the very applicability of the Gibbs statistical mechanics, where all parameters must be small compared to the number of particles N (i.e. the first limit must be infinite N, not v. Any alternative is meaningless: to justify v>>N, the larger binding energy must grossly exceed the nuclear energy even for very moderate N). In such singular case, the component with infinite binding energy divides a random heteropolymer into random length homopolymers with bounded edges, which exclude their interaction. This replaces the unbinding of a disordered heteropolymer in Eq. (1) in [1] with a different problem of unbinding in an ensemble of random length homopolymers in Eq. (2). Since only an infinite homopolymers length L implies singularity, and since the probability of such length is exponentially small in N, the ensemble unbinding may (but must not) yield only a Griffiths singularity. Accurate calculation [from Eq. (2) to Eq. (16)] verifies this statement by proving the continuity of all finite derivatives. However, even in the considered case calculation neither specifies the Griffiths singularity (which is a generic name for any

"exponentially weak" singularity) analytically, nor proves the existence of any singularity.  The only "insight of the singularity" is provided by Eq. (17), which yields the second order phase transition. The latter is in sharp contrast to Griffiths singularity, and demonstrates only the invalidity of the corresponding approximation. The final KM speculation "The fact that v is finite but large is not expected to modify the nature of the singularity" is certainly not a substitute for an accurate generalization of the special case.

The same authors and Peliti (KMP, [2]) analytically accounted for the excluded-volume interactions between unbounded loops and the rest of the polymer to calculated the  loop entropy [2]. In a homopolymer such entropy yields the first order phase transition [3]. The authors "comment on disorder" and argue: "Analysis strongly suggests that the transition is first order… in agreement with experiments, [where] sharpness of the jumps [in the DNA unbinding curve] suggests that the transition from bound to unbound is first order".  (The authors do not specify whether and how it is related to weak disorder). In a heteropolymer both suggestions are invalid. Contrary to the second suggestion, multiple, progressively (with increasing temperature T) sharper and higher, peaks imply stepwise unbinding of heterogeneous DNA. The length of progressively refractory bound domains increases, and ultimately diverges; the length of unbound domains between them is exponentially larger [4]. (This crucial implication of heterogeneity is disregarded in [2]). Experimental peaks [5] quantitatively agree with analytical theory [4] and allow for DNA sequencing [6]. (Not by chance, the titles of the Wartell, Benight and Gotoh papers [4] are "Thermal denaturation of DNA molecules-a comparison of theory with experiments" and "Prediction of melting profiles and local helix stability for sequenced DNA").   In sharp contrast to the first suggestion, accurate theory [7], which properly accounts for

DNA heterogeneity), yields a high (~100) order phase transition in the extraordinary narrow vicinity of the transition temperature. At lower temperatures the transition is preceded with the specified Griffiths type dependence. Such singularity is slightly smoother than the Kosterlitz -Thouless singularity in the case of strong disorder [8]. However, the difference may be related to directed heteropolymers considered in [8], and to possibly insufficient accuracy of their renormalization approach in the extraordinary narrow region of the high order transition.

To summarize. Contrary to KMP [2], the transition in a disordered heteropolymer is never first order. KM [1] established (but not specified analytically) the Griffiths singularity in an ensemble of random length homopolymers only.

Mark Ya. Azbel'

School of Physics and Astronomy, Tel Aviv University, Tel Aviv 69978 Israel

References

1. Y. Kafri and D. Mukamel, Phys. Rev. Lett. **91**, 055502-1 (2003)

2. Y. Kafri, D. Mukamel, and L. Peliti, Phys. Rev. Lett. **85**, 4988 (2000)

3. M.E. Fisher, J. Chem. Phys. **45**, 1469 (1966)

4. M. Ya. Azbel', Phys. Rev. A **20**, 1671 (1979).

5. A. S. Benight, R. M. Wartell, D. K. Howell, Nature **289**, 203 (1981); O. Gotoh, Adv. Biophys. **16**, 1 (1983); R. M. Wartell, A. S. Benight, Phys. Rep. **126**, 67 (1985)

6. M. Ya. Azbel', Proc. Nat. Acad. USA **76**, 101 (1979); Biopolymers **19**, 61, 81, 95, 1311 (1980)

7. M. Ya. Azbel', Phys. Rev. E **68**, 050901 (R) (2002)

8. L.-H. Tang and M. Chate, Phys. Rev. Lett. **86**, 830 (2001)